\documentclass[
    ,final            
  ]
  {aipproc}

\layoutstyle{8x11single}

\usepackage{amssymb}

\begin{document}

\title{Mesoscopic superconductivity in ultrasmall metallic grains}


\author{Y. Alhassid}{
  address={Center for Theoretical Physics, Sloane Physics Laboratory, Yale University, New Haven, CT 06520, USA}  
}

\author{K.  N. Nesterov}{
  address={Center for Theoretical Physics, Sloane Physics Laboratory, Yale University, New Haven, CT 06520, USA}
  ,altaddress={Current address: Univ. Grenoble Alpes, INAC-SPSMS, F-38000 Grenoble, France
and CEA, INAC-SPSMS, F-38000 Grenoble, France } 
}

\begin{abstract}
A nano-scale metallic grain (nanoparticle) with irregular boundaries in which the single-particle dynamics are chaotic is a zero-dimensional system described by the so-called universal Hamiltonian in the limit of a large number of electrons. The interaction part of this Hamiltonian includes a superconducting pairing term and a ferromagnetic exchange term. Spin-orbit scattering breaks spin symmetry and suppresses the exchange interaction term. Of particular interest is the fluctuation-dominated regime, typical of the smallest grains in the experiments, in which the bulk pairing gap is comparable to or smaller than the single-particle mean-level spacing,  and the Bardeen-Cooper-Schrieffer (BCS) mean-field theory of superconductivity is no longer valid. 
Here we study the crossover between the BCS and fluctuation-dominated regimes in two limits.
In the absence of spin-orbit scattering, the pairing and exchange interaction terms compete with  each other. We describe the signatures of this competition in thermodynamic observables, the heat capacity and spin susceptibility. In the presence of strong spin-orbit scattering, the exchange interaction term can be ignored. We discuss how the magnetic-field response of discrete energy levels in such a nanoparticle is affected by pairing correlations. We identify signatures of pairing correlations in this response, which are detectable even in the fluctuation-dominated regime. 
\end{abstract}

\maketitle


\section{Introduction}

In a bulk metal, the single-electron spectrum can be considered quasi-continuous. However, in ultra-small grains, the discreteness of the single-electron spectrum becomes important and is characterized by its mean level spacing $\delta$. In a sufficiently small superconducting grain, $\delta$ may be comparable to or larger than the bulk pairing gap $\Delta$. This has been achieved in single-electron-tunneling spectroscopy experiments on metallic nanoparticles~\cite{vonDelft2001}, where the crossover between the regimes with $\Delta/\delta >1$ (larger grains) and $\Delta/\delta <1$ (smaller grains) was observed in Al grains~\cite{Black1996}. 

The regime $\Delta/\delta \lesssim 1$, referred to as the fluctuation-dominated regime, is of special interest. 
According to Anderson's criterion~\cite{Anderson1959_jpcs}, superconductivity is no longer possible in this regime.  However, this does not necessarily  imply the complete suppression of pairing correlations.  Here we discuss two possible probes to detect signatures of pairing correlations: (i) thermodynamic observables, and (ii) for a grain with strong spin-orbit scattering, the response of discrete energy levels to an external magnetic field.

The Bardeen-Cooper-Schrieffer (BCS) theory of superconductivity~\cite{Bardeen1957} describes well the properties of a superconducting grain in the limit $\Delta/\delta \gg 1$, but breaks down when $\Delta/\delta \lesssim 1$. To describe this regime, it is then necessary to take into account the fluctuations of the order parameter.  In nuclei, typically $\Delta/\delta \sim 3$, and thus the crossover regime between the BCS and fluctuation-dominated regime is of common interest to both the mesoscopic and nuclear physics communities~\cite{Alhassid2013_book}. 

A nano-scale metallic grain with irregular boundaries, in which the single-electron dynamics are chaotic, is described by the so-called universal Hamiltonian for  a large number of electrons~\cite{Kurland2000,Aleiner2002}. The one-body part of the universal Hamiltonian is described by random-matrix theory~\cite{Mehta1991,Alhassid2000} and its interaction part includes a pairing term and a ferromagnetic spin-exchange term. For a fixed number of electrons  and in the absence of spin-orbit scattering and orbital magnetic field, it has the form  
\begin{equation}\label{universal_hamiltonian}
 \hat{H} = \sum_{k\sigma} \epsilon_k c^\dagger_{k\sigma} c_{k\sigma} - G\hat{P}^\dagger \hat{P} - J_s \hat{\textbf{S}}^2\,,
\end{equation}
where
\begin{equation}
 \hat{P} = \sum_k c_{k\downarrow}c_{k\uparrow}\,,\quad \hat{P}^\dagger = \sum_k c^\dagger_{k\uparrow} c^\dagger_{k\downarrow}\,,
\end{equation}
and $\textbf{S}$ is the total spin of electrons in the grain. The pairing and exchange terms compete with each other. Pairing interaction tends to minimize the spin of the nanoparticle while exchange interaction tends to maximize spin polarization. The competition between pairing and exchange correlations leads to a coexistence regime of superconductivity and ferromagnetism in the fluctuation-dominated regime~\cite{Ying2006,Schmidt2007,Alhassid2012}. Here we review signatures of this competition in thermodynamic properties of the nanoparticle such as the heat capacity and spin susceptibility~\cite{VanHoucke2010,Alhassid2012, Nesterov2013}. Number-parity effects, induced by pairing correlations~\cite{Muhlschlegel1972, DiLorenzo2000, Falci2000, Schechter2001, Falci2002, Gladilin2004,VanHoucke2006}, are modified by the exchange interaction in a way that is qualitatively different between the fluctuation-dominated and BCS regimes. This is discussed below, and more details can be found in Refs.~\cite{Alhassid2012}
 and \cite{Nesterov2013}.

The addition of spin-orbit scattering to the single-particle part of the Hamiltonian (\ref{universal_hamiltonian}) breaks spin-rotation symmetry but preserves time-reversal symmetry. Consequently, the exchange interaction is suppressed while the pairing term remains unaffected. We studied the response of discrete energy levels of a nanoparticle with strong spin-orbit scattering to an external magnetic field. In particular, we studied the linear and quadratic terms in the magnetic field that are parametrized by the $g$-factor and level curvature, respectively. We find that, while the $g$-factor statistics are unaffected by pairing correlations, the level curvature statistics are highly sensitive to them and can thus be used to probe pairing correlations in the single-electron-tunneling spectroscopy experiments. We discuss this below, and more details will be published elsewhere~\cite{Nesterov2014}.

\section{Superconducting nanoparticles in the absence of spin-orbit scattering: thermodynamic observables}

\begin{figure}[h!]
 \includegraphics[width=0.65\textwidth]{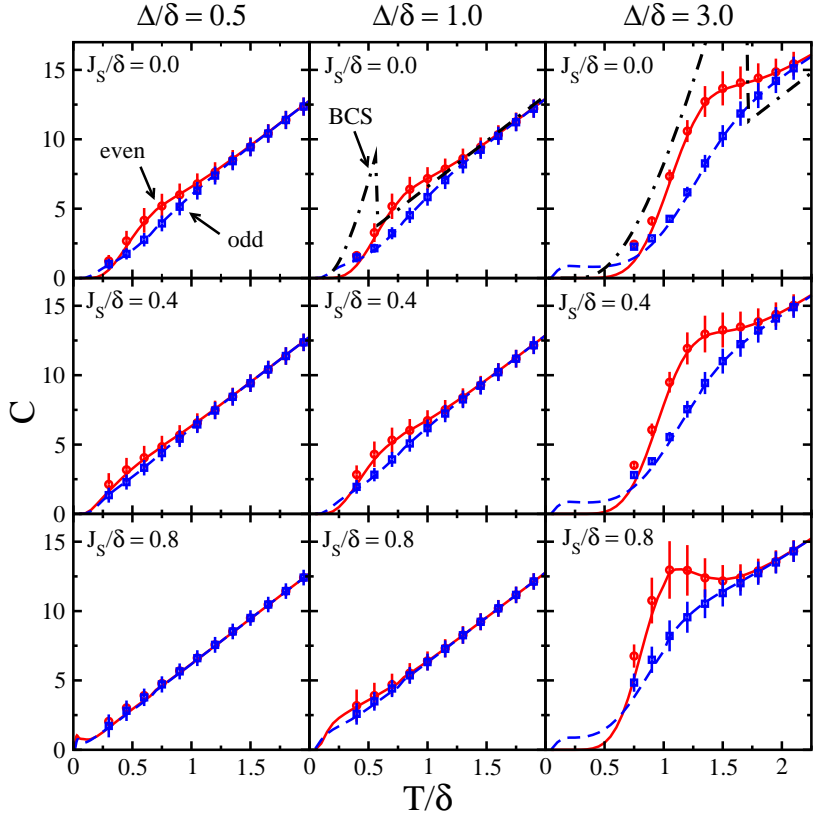}\caption{The heat capacity versus temperature $T/\delta$ for  even (solid lines, circles) and odd (dashed lines, squares) grains. Results are shown for grains with $\Delta/\delta=0.5$ (left column), $\Delta/\delta=1.0$ (middle column), and $\Delta/\delta=3.0$ (right column), and for $J_s/\delta=0.0$ (top row), $J_s/\delta = 0.4$ (middle row), and for $J_s/\delta = 0.8$ (bottom row). The averages over the mesoscopic fluctuations are shown by symbols, and the standard deviations are described by vertical bars. The lines are the results for an equally spaced single-particle spectrum in the Hamiltonian (\ref{universal_hamiltonian}). The solid and dashed lines are obtained by using Richardson's solution at low temperature and the SPA+RPA approach at higher temperature, and the dash-dotted lines are the results of the grand-canonical BCS calculations (where applicable). Taken from Ref.~\cite{Nesterov2013}.}\label{Fig_hc}
\end{figure}

\begin{figure}[h!]
 \includegraphics[width=0.65\textwidth]{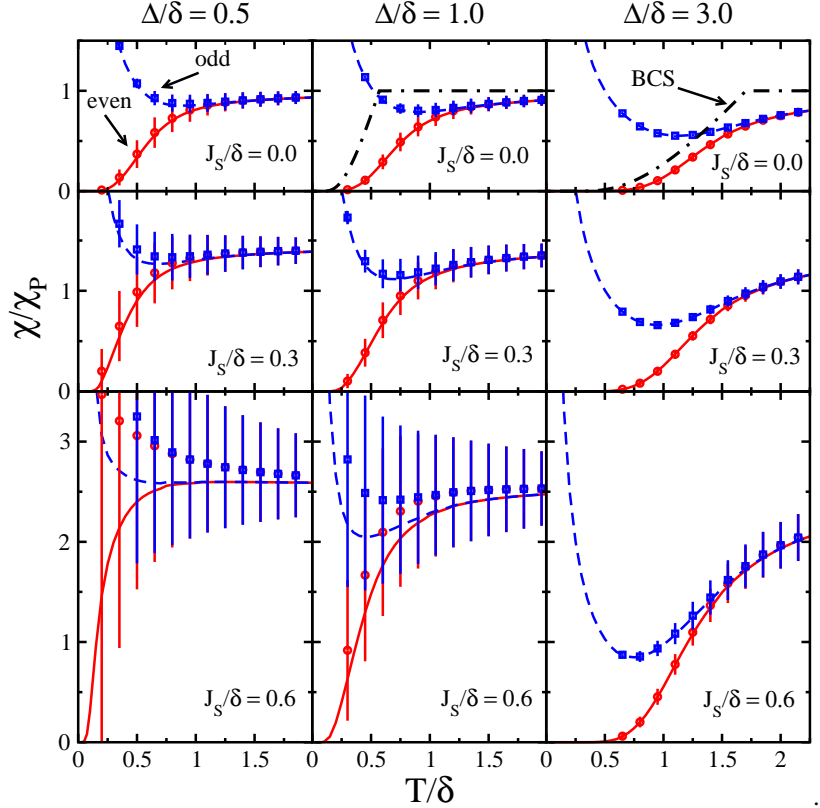}\caption{The spin susceptibility $\chi$ in the units of $\chi_P = 2\mu_B^2/\delta$ versus temperature $T/\delta$ for even and odd grains for the same values of $\Delta/\delta$ as in Fig.~\ref{Fig_hc}, and for $J_s/\delta = 0.0$ (top row), $J_s/\delta = 0.3$ (middle row), and $J_s/\delta = 0.6$ (bottom row). The notation follows the same convention as in Fig.~\ref{Fig_hc}.  Taken from Ref.~\cite{Nesterov2013}. }\label{Fig_ss}.
\end{figure}

In a small superconducting grain, the heat capacity and spin susceptibility display odd-effect effects, i.e., they are sensitive to the parity of the number of electrons in the grain. Examples are shown in the top panels of Figs.~\ref{Fig_hc} and \ref{Fig_ss} for various values of $\Delta/\delta$ (solid lines correspond to even particle number and dashed lines to odd particle number). 
The odd-even effect in the heat capacity at temperatures $\sim \max(\delta,\Delta)$ is characterized by the enhancement of the heat capacity for an even grain in comparison with an odd grain. Similar odd-even effects in the heat capacity of nuclei were observed in experiments~\cite{Schiller2001} and in microscopic calculations~\cite{Liu2001}.  As $\Delta/\delta$ increases, this enhancement transforms into a shoulder, which in the BCS limit $\Delta/\delta \gg 1$ becomes a discontinuity at the BCS critical temperature.   
At low temperatures $T \ll \max(\delta,\Delta)$, the effect is characterized by an exponential suppression of the even-grain heat capacity, caused by the excitation gap. Another interesting observable is the spin susceptibility. Pairing correlations tend to suppress it, and the even-grain spin susceptibility is exponentially suppressed at low temperatures. For odd number of electrons in the grain, there is an unpaired spin that leads to a Curie-like $1/T$ divergence at low temperatures.   The combination of this Curie-like behavior with a competing suppression caused by pairing  correlations results in a local minimum of the spin susceptibility. This so-called re-entrant behavior is a unique signature of pairing correlations in the spin susceptibility~\cite{DiLorenzo2000,Falci2000}.  Similar effects were found in the moment of inertia of nuclei~\cite{Alhassid2005_prc}, which measures their response to rotations.

It is interesting to determine how the number-parity signatures of pairing correlations are affected by the exchange interaction. Mesoscopic fluctuations of the thermodynamic observables must also be taken into account. For example, it was found that the exponential suppression of the normal-grain heat capacity at low temperatures for an equally spaced spectrum is replaced by power law when fluctuations of single-particle energy levels are taken into consideration and the heat capacity is averaged over them~\cite{Denton1971,Denton1973}.  

We have employed an efficient and semi-analytic technique to calculate thermodynamic observables in a grain that is described by the universal Hamiltonian. In the following, we discuss the main ingredients of this method and more details can be found in Ref.~\cite{Nesterov2013}. The exchange interaction is treated exactly using a spin projection technique. Rewriting the universal Hamiltonian in Eq.~(\ref{universal_hamiltonian}) as $\hat H = \hat{H}_{\mathrm{BCS}} - J_s \hat{\textbf{S}}^2$,  its partition function is given by
\begin{equation}\label{partition}
 \mathrm{Tr} e^{-\beta(\hat{H}_{\mathrm{BCS}} - J_s \hat{\textbf{{S}}}^2)} = \sum_S e^{\beta J_s S(S+1)} \mathrm{Tr}_S e^{-\beta \hat{H}_{\mathrm{BCS}}}\,,
\end{equation}
where we have used $[\hat{H}_{\mathrm{BCS}}, \hat{\textbf{{S}}}]=0$. Here, $\mathrm{Tr}_S$ denotes the trace of an operator taken with respect to the subspace of many-body states with a fixed value $S$ of the total-spin quantum number. This trace can be evaluated  using spin projection, which in general requires a three-dimensional integral over the Euler angles~\cite{Ring1980}. However, for a scalar operator $\hat X$, the quantity
 $\mathrm{Tr}_S \hat{X}$ can be simply related to quantities of the form $\mathrm{Tr}_{S_z=M} \hat{X}$, whose evaluation amounts to projection on fixed values of the spin component $S_z=M$~\cite{Alhassid2007_prl}
\begin{equation}\label{spin-projection}
 \mathrm{Tr}_S \hat{X} = (2S+1) (\mathrm{Tr}_{S_z = S} \hat{X} - \mathrm{Tr}_{S_z=S+1} \hat{X})\,.
\end{equation}
Relations (\ref{partition}) and (\ref{spin-projection}) enable us to express the partition function of a grain described by the universal Hamiltonian (\ref{universal_hamiltonian}) to the $S_z$-projected partition function of a grain described by the BCS-like Hamiltonian $\hat{H}_{\mathrm{BCS}}$.

The pairing Hamiltonian $\hat{H}_{\mathrm{BCS}}$ is treated using a path-integral formalism. In this approach, the imaginary-time propagator is written as a path integral over a complex $\tau$-dependent auxiliary field $\widetilde{\Delta}$
\begin{equation}
 \exp \left[-\beta\left(\hat{H}_{\mathrm{BCS}} - \mu\hat{N}\right)\right] = \int \mathcal{D}[\widetilde{\Delta},\widetilde{\Delta}^*] \mathcal{T} \exp \left[- \int\limits_0^\beta d\tau \left( \frac{|\widetilde{\Delta}(\tau)|^2}{G} + \hat{H}_{\widetilde{\Delta}(\tau)}\right)\right]\,,
\end{equation}
where
\begin{equation}
 \hat{H}_{\widetilde{\Delta}} = \sum_k \left[ \left(\epsilon_k - \mu- \frac G2\right)(c^\dagger_{k\downarrow} c_{k\downarrow} + c^\dagger_{k\uparrow}c_{k\uparrow}) - \widetilde{\Delta} c^\dagger_{k\uparrow} c^\dagger_{k\downarrow} - \widetilde{\Delta}^* c_{k\downarrow} c_{k\uparrow} + \frac G2\right]
\end{equation}
is a one-body Hamiltonian describing noninteracting electrons in an external pairing field, $\mu$ is the chemical potential, and $\mathcal{T}$ denotes the time-ordering. The saddle-point value of $\widetilde{\Delta}$ in the grand-canonical formalism results in the BCS theory. We decompose the field into its time-independent (static) part $\widetilde{\Delta}_0 = (1/\beta) \int \limits_0^\beta \widetilde{\Delta}(\tau) d\tau$ and its fluctuating time-dependent part $\delta \widetilde{\Delta}(\tau) = \widetilde{\Delta}(\tau) - \widetilde{\Delta}_0$. We evaluate the integral over $\widetilde{\Delta}_0$ exactly, which alone would result in the static-path approximation (SPA)~\cite{Muhlschlegel1972, Alhassid1984, Lauritzen1988}, and treat the remaining integral over $\delta \widetilde{\Delta}(\tau)$ in the saddle-point approximation around each static $\widetilde{\Delta}_0$. The latter leads to a random-phase-approximation-like (RPA-like) correction to the SPA result for each $\widetilde{\Delta}_0$. This technique 
is known as the SPA+RPA approach~\cite{Kerman1981, Puddu1991, Lauritzen1993, Rossignoli1997, Attias1997}.

The number of electrons in the grain is essentially fixed by the large charging energy, leading to the odd-even effects  discussed above.  Canonical ensemble calculation at a fixed number of electrons results in cumbersome expressions. Instead, we work in the grand-canonical formalism, but account for odd-even effects using the  number-parity projection~\cite{Goodman1981,Rossignoli1998,Balian1999,Falci2002}
\begin{equation}
 \hat{P}_\eta = \frac 12 \left( 1+ \eta e^{i\pi \hat{N}}\right)\,.
\end{equation}
With such projection, only many-electron states with even (odd) particle number are taken into account for $\eta=+1$ ( $\eta =-1$). Particle-number fluctuations in the grand-canonical formalism are accounted for by performing the canonical projection in the saddle-point approximation (which alone cannot describe the number-parity effects). 

\begin{figure}
 \includegraphics[width=0.7\textwidth]{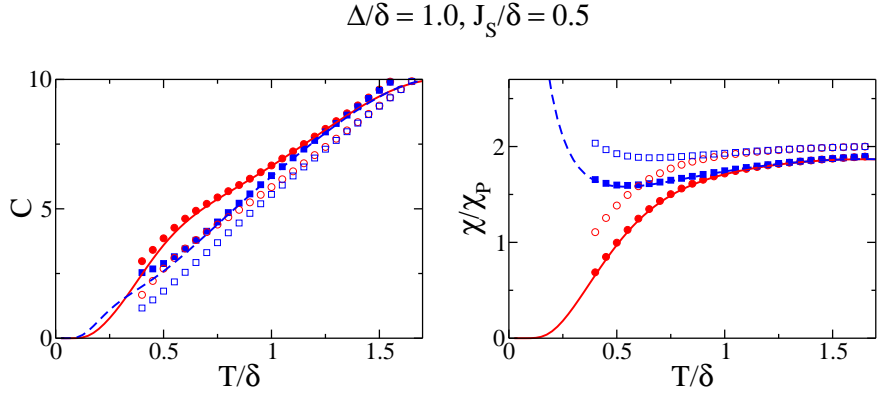}\caption{The heat capacity (left) and spin susceptibility (right) calculated for a specific realization of the single-particle RMT spectrum in the universal Hamiltonian (\ref{universal_hamiltonian}) for a grain with $\Delta/\delta = 1.0$ and $J_s/\delta = 0.5$. The exact canonical results are shown by lines, the results of the number-parity projected SPA  are shown by open symbols, and the  SPA+RPA calculations are shown by the solid symbols. The even grain results are shown by solid lines and circles, while the odd grain results are described by dashed lines and squares. The spin susceptibility is measured in the units of the Pauli susceptibility $\chi_P = 2\mu_B^2/\delta$. Adapted from Ref.~\cite{Alhassid2012}.}\label{Fig_rich}
\end{figure}

The methods discussed above are applied for each random-matrix realization of the single-particle spectrum in the Hamiltonian~(\ref{universal_hamiltonian}). To study the mesoscopic fluctuations of observables, we generate many such realizations, sampled from the Gaussian orthogonal ensemble (GOE) of RMT. We note that, in principle, for each realization of the single-particle spectrum, many-body calculations can be performed exactly using Richardson's solution~\cite{Richardson1963,Richardson1967,Schmidt2007} or quantum Monte-Carlo simulations~\cite{VanHoucke2010}. However, both methods are impractical when many realizations of the grain have to be studied. In addition, Richardson's solution becomes prohibitively difficult for large values of $\Delta/\delta$ or at high temperatures, when the number of required many-electron eigenvalues grows exponentially. In Fig.~\ref{Fig_rich} we demonstrate the validity of our method for particular realizations of the single-particle spectrum  by comparing with Richardson's 
solution. We observe that the number-parity-projected SPA+RPA calculations agree well with the 
exact results, while the SPA calculations are insufficient. We note that the SPA+RPA method cannot be used at very low temperatures, when the Gaussian fluctuations around certain static values of the field become unstable.~\footnote{This problem can be overcome by identifying and treating nonperturbatively a low-energy collective mode~\cite{Ribeiro2012}.}

The main results of our calculations are shown in Fig.~\ref{Fig_hc} (heat capacity) and in Fig.~\ref{Fig_ss} (spin susceptibility) and can be summarized as follows:
\begin{itemize}
 \item The exchange interaction can shift the odd-even effect to lower temperatures and can completely destroy certain signatures of pairing correlations. For example, there is no re-entrant behavior for the average spin susceptibility in grains with $\Delta/\delta=0.5$ and $J_s/\delta = 0.6$. 
 \item The effects of the exchange interaction are qualitatively different in the fluctuation-dominated and BCS regimes. While in the fluctuation-dominated regime the main effect of exchange correlations is to suppress the signatures of pairing correlations, certain signatures are enhanced when $\Delta/\delta > 1$. For example, in a grain with $\Delta/\delta = 3.0$,  the exchange interaction transforms the shoulder in the even-grain heat capacity into a peak, and the re-entrant behavior is enhanced. 
 \item In the fluctuation-dominated regime, the mesoscopic fluctuations can completely destroy odd-even effects, when the standard deviations of thermodynamic observables in the even and odd grains overlap. 
  \item The fluctuations of the spin susceptibility become especially strong in the presence of exchange in the fluctuation-dominated regime since the spin susceptibility is highly sensitive to excitation energies of states with larger spins. 
\end{itemize}

In practice, $J_s/\delta$ is a material-dependent parameter, while $\Delta/\delta$ can be tuned by changing the size of a grain. Thus, for a given material, one can only tune the value of $\Delta/\delta$. Since the exchange is much more effective in suppressing the odd-even effects in the fluctuation-dominated regime than in the larger grains, we conclude that, for a given material, exchange interaction effectively controls how fast the odd-even effects get suppressed when the size of the grain is decreased. For a material with small $J_s/\delta$, the odd-even effects may survive down to the smallest grains, while in the presence of stronger exchange, the odd-even effects are destroyed in the smaller grains.

\section{Superconducting nanoparticles with spin-orbit scattering: magnetic-field response of discrete energy levels}

Spin-orbit scattering breaks spin-rotation symmetry and suppresses the exchange term in the universal Hamiltonian (\ref{universal_hamiltonian}). When it is sufficiently strong, the relevant ensemble for the one-body part of the Hamiltonian is the Gaussian symplectic ensemble (GSE).  Since the pairing interaction is not suppressed by spin-orbit scattering, the universal Hamiltonian has the form
\begin{equation}\label{spin-orbit-H}
 \hat{H} = \sum_{k\alpha} \epsilon_k c^{\dagger}_{k \alpha} c_{k\alpha} - G\hat{P}^\dagger \hat{P}
\end{equation}
with
\begin{equation}\label{pair}
 \hat{P} = \sum_k c_{k2}c_{k1}\,,\quad \hat{P}^\dagger = \sum_k c^\dagger_{k1} c^\dagger_{k2} \;.
\end{equation}
In the Hamiltonian (\ref{spin-orbit-H}), spin is no longer a good quantum number. However, since time-reversal symmetry is preserved, the single-particle eigenstates come in doubly degenerate pairs, and the states in each pair are related by time-reversal symmetry. The label $\alpha=1,2$ in Eq.~(\ref{pair})  denotes a pair of such Kramers-degenerate states for a given $k$. 

\begin{figure}
 \includegraphics[width=0.7\textwidth]{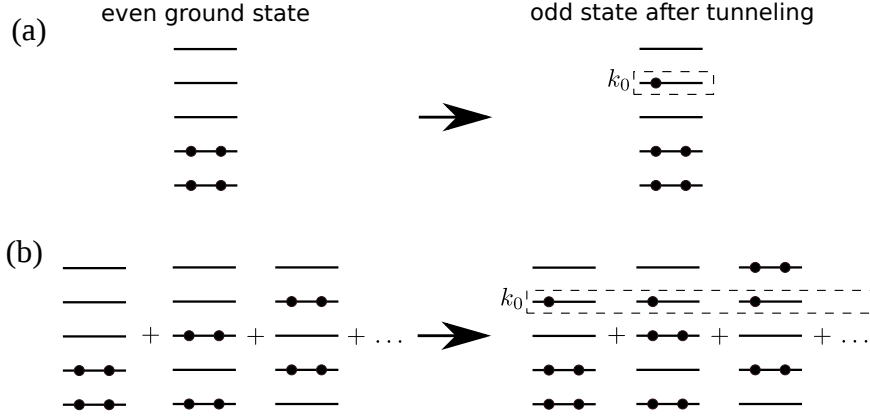}\caption{Many-electron states participating in a process of tunneling onto an even grain in its ground state in the absence (top panels) and in the presence (bottom panels) of pairing correlations.}\label{Fig_states}
\end{figure}

In the absence of pairing correlations, the energies extracted in a single-electron-tunneling spectroscopy experiment correspond to the single-particle energies. For example, when the initial state in a process of electron tunneling into the grain is the ground state of the grain with an even number of electrons, an electron can only tunnel into one of the previously empty orbitals above the Fermi level, as is  shown on the top panels of Fig.~\ref{Fig_states}. The quantity extracted experimentally is the difference between the many-body energies of the two states, and is equal to the single-particle energy of the singly occupied orbital in the odd state [up to a particle-number-dependent constant that is due to the charging energy omitted in Eq.~(\ref{spin-orbit-H})]. 
These single-particle energies split in a magnetic field; for a weak magnetic field, this splitting can be parametrized by $g$-factors and level curvatures (at zero magnetic field) defined by
\begin{equation}\label{definition_response_sp}
 \epsilon_k (B) = \epsilon_k (0) \pm \frac 12 g_k \mu_B B + \frac 12 \kappa B^2  + O(B^3)\,.
\end{equation}
In the absence of spin-orbit scattering, all $g$-factors are determined by the electron spin and are simply equal to 2.  Spin-orbit scattering leads to level-to-level fluctuations of $g$-factors and to their dependence on the magnetic-field direction. These effects were studied extensively using RMT~\cite{Matveev2000,Brouwer2000}, and the theoretical predictions for the $g$-factor distribution agree well with experiments in noble-metal nanoparticles~\cite{Petta2001,Petta2002,Kuemmeth2008}. In one of these experiments~\cite{Kuemmeth2008}, the level curvature distribution was also measured and found to agree with RMT predictions~\cite{Fyodorov1995,vonOppen1995}. 

Here we are interested in determining the effects of pairing correlations on the  $g$-factor and level curvature statistics, i.e., we consider superconducting nanoparticles with spin-orbit scattering. To proceed, we first generalize the single-particle definition (\ref{definition_response_sp}) of these quantities to the interacting case. We discuss the special case of asymmetric tunnel barriers (one-bottleneck geometry), when the current is determined by the processes of tunneling onto the grain in its ground state~\cite{vonDelft2001}. We consider the conductance peak that corresponds to the tunneling onto the grain with an even number of electrons. Then, we define the $g$-factor and level curvature in analogy with Eq.~(\ref{definition_response_sp}), but for the extracted difference  between the corresponding many-electron energies
\begin{equation}\label{definition_response_mb}
 E^{N_e+1}_{\Omega}(B) - E^{N_e}_0(B) =  E^{N_e+1}_{\Omega}(0) - E^{N_e}_0(0) \pm \frac 12 g \mu_B B + \frac 12 \kappa B^2 + O(B^3)\,.
\end{equation}
Here $E^{N_e}_0$ is the energy of the even ground state $|0\rangle$ with $N_e$ electrons and $E^{N_e+1}_{\Omega}$ is the energy of a possible final odd state $|\Omega\rangle$ with $N_e+1$ electrons. The splitting of this difference in a magnetic field occurs because of the splitting of the energy of the doubly degenerate final odd state. 

In the following,  we summarize our main findings; additional details will be published elsewhere~\cite{Nesterov2014}. 
The $g$-factor is determined by corrections to the eigenenergies of the two states participating in a tunneling process that are linear in the external magnetic field.
In the presence of pairing, the even ground state is a superposition of Slater determinants with doubly occupied levels, as illustrated in the left panel of Fig.~\ref{Fig_states}(b). This many-particle state is invariant under time reversal and is thus non-degenerate. Since the magnetization itself is odd under time reversal, we conclude that its expectation value in the even ground state is identically zero. 

In a possible odd state $|\Omega\rangle$, one of the single-particle orbitals is singly occupied. Since pairing correlations can only scatter a pair of electrons  from a doubly occupied level to an empty level, the singly occupied orbital remains the same in all the terms of the superposition [as is shown in the right panel of Fig.~\ref{Fig_states}(b)]. In nuclear physics this is known as the blocking effect of the pairing interaction~\cite{Soloviev1961}. Because of this blocking effect, the matrix element of the magnetization operator between two states of the same many-particle doublet separates into a contribution from the single electron on the blocked orbital and the contribution from the remaining fully-paired electrons. The latter contribution is identically zero following the time-reversal symmetry considerations used for the even ground state. The remaining single-electron contribution is just the corresponding single-particle matrix element. We conclude that the many-body $g$-factor defined in Eq.~(
\ref{definition_response_mb}) is exactly 
the single- particle $g$-factor defined in Eq.~(\ref{definition_response_sp}) for the blocked orbital in the odd state.  Thus, the  $g$-factor distribution is unaffected by pairing correlations. 

We emphasize that this conclusion regarding the $g$-factor statistics is very general. It is independent of the particular statistics of the single-electron energies or of the relative contributions of orbital and spin magnetism to the splitting of energy levels. Since the exchange interaction (which is not fully suppressed when spin-orbit is not strong) affects $g$-factors statistics~\cite{Gorokhov2003,Gorokhov2004}, we conclude that the robustness of the $g$-factor statistic can be a good test to determine the importance of exchange correlations in the grain. 

In contrast, we find that level curvatures are highly sensitive to pairing correlations. Here we only consider the transition between the even and odd ground states (that corresponds to the first conductance peak). 
In the absence of pairing correlations, the level curvature reduces to the single-particle level curvature
\begin{equation}
 \kappa_k = 2\sum_{k'\ne k} \frac{|\langle k1|\hat{M}_z| k'1\rangle|^2 + |\langle k1|\hat{M}_z| k'2\rangle|^2}{\epsilon_k - \epsilon_{k'}}\,,
\end{equation}
where $\langle k\alpha |\hat{M}_z| k'\alpha'\rangle$ is the single-particle matrix element of the component of the magnetization operator $\hat M_z$ along the $z$ axis (chosen to be the direction of the magnetic field).  It is apparent that the average of such a quantity is zero and that its distribution is symmetric. 

In the presence of interactions,  the level curvature defined in Eq.~(\ref{definition_response_mb}) is given by 
\begin{equation}\label{kappa_int}
 \kappa = 2\sum'_{\Omega'} \frac{\left|\langle 0|\hat{M}_z|\Omega'\rangle \right|^2}{E^{N_e+1}_0 - E^{N_e+1}_{\Omega'}} - 2\sum'_{\Theta'} \frac{\left|\langle 0|\hat{M}_z|\Theta'\rangle \right|^2}{E^{N_e}_0 - E^{N_e}_{\Theta'}}\,,
\end{equation}
where the sums are taken over excited $N_e+1$- and $N_e$-particle states $|\Omega'\rangle$ and $|\Theta'\rangle$, respectively. Since the second-order correction to the ground-state energy is always negative, we observe that the contributions to $\kappa$ from the odd-grain states are always negative, while the contributions from the even-grain states are always positive. Similarly, level curvatures in the right tail of the distribution correspond to even grains with small excitation energy, while level curvatures in the left tail of the distribution arise from odd grains with small excitation energy. 

In the presence of pairing correlations, the excitation spectrum of the even grain develops a gap, which on average suppresses the contribution from the even states and, in particular, suppresses the right tail of the distribution. The excitation spectrum of the odd grain does not develop such a gap and the density of low-lying excitations is enhanced by pairing. Therefore, the negative contribution to curvatures is enhanced by pairing correlations and the net curvature becomes negative on average, while the resulting distribution becomes asymmetric. 

\begin{figure}
 \includegraphics[width=\textwidth]{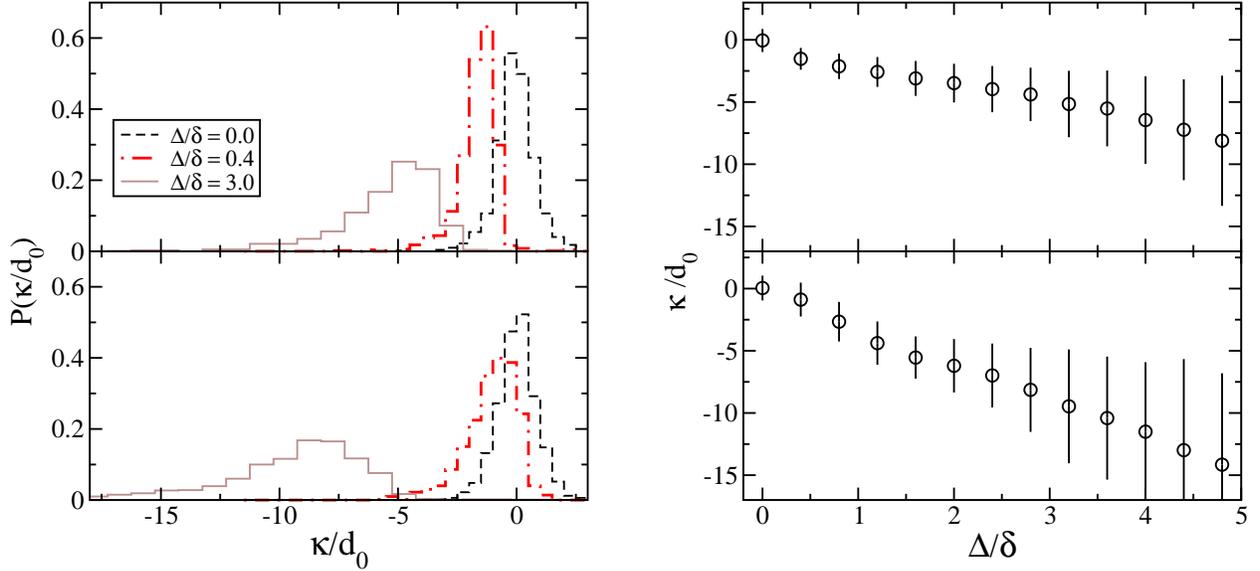}\caption{Level curvatures of the first conductance peak (corresponding to the transition between the even and odd ground states). Left panels: level curvature distributions  for $\Delta/\delta=0$, $\Delta/\delta=0.4$ and $\Delta/\delta = 3.0$.   Right panels: the median curvature versus $\Delta/\delta$. The top panels are calculated using exact diagonalization of the many-body Hamiltonian in a model space of 13 degenerate single-particle levels. The bottom panels are calculated using a BCS-like approach. The error bars of the median curvatures  denote the midspread (i.e., the middle 50\%) of the distribution. The level curvatures are expressed in units of the midspread $d_0$ of the single-particle curvature distribution so that the dependence on the random-matrix dimension is eliminated.}\label{Fig_curv}
\end{figure}

We have verified our qualitative arguments by performing exact numerical calculations as well as calculations based on a BCS-like formalism that takes into account the blocking effect. In both calculations, we have neglected the orbital contribution to $\hat{M}_z$, i.e., assumed that $\hat{M}_z = 2\mu_B\hat{S}_z$, and considered the GSE statistics for the single-particle states in Eq.~(\ref{spin-orbit-H}).  In the exact calculations, we work in a relatively small model space of 13 single-particle orbitals since the size of the many-electron model space grows combinatorially with the number of single-particle orbitals. In the BCS calculations, the single-particle model space included 121 levels. 

We present our main results in Fig.~\ref{Fig_curv}.  We find that in the presence of pairing correlations, the level curvature distribution becomes asymmetric and is shifted to negative values. In addition, its dispersion grows significantly with increasing $\Delta/\delta$. This can be explained by an increasing probability to have small denominators for the odd-grain contribution in Eq.~(\ref{kappa_int}) as the density of odd excitations grows.  It is remarkable that these effects are already appreciable in the fluctuation-dominated regime. Thus level curvatures offer a sensitive probe  of pairing correlations in this fluctuation-dominated regime.  

\section{Conclusion}

In conclusion, we have discussed the signatures of pairing correlations in ultrasmall superconducting nanoparticles both in the absence and in the presence of spin-orbit scattering. Grains with chaotic single-electron dynamics are generally described by the universal Hamiltonian. In the absence of spin-orbit scattering, this Hamiltonian includes competing pairing and exchange correlations. Pairing correlations  lead to number-parity effects in thermodynamic observables of such grains. In the smaller grains, the exchange interaction can suppress and even completely destroy these odd-even effects. However, in larger grains, it can enhance certain signatures of pairing correlations such as the re-entrant behavior of the spin susceptibility. When mesoscopic fluctuations  are taken into account, the odd-even effects in thermodynamic observables may be washed out, and the fluctuations of the spin susceptibility may be especially large. 

In superconducting grains with strong spin-orbit scattering, exchange correlations are suppressed. We have studied how pairing correlations affect the response of discrete energy levels of the grain to an external magnetic field. We have found that the distribution of $g$-factors (characterizing the linear correction in the magnetic field) is not affected by pairing correlations, while the distribution of level curvatures  (the second-order correction) is highly sensitive to them. Since level curvatures can be measured in single-electron-tunneling spectroscopy experiments, they offer a sensitive probe to detect pairing correlations. This probe is particularly useful in the fluctuation-dominated regime.

\begin{theacknowledgments}
This work was supported in part by the U.S. DOE grant No. DE-FG02-91ER40608. Computational cycles were provided by the facilities of the Yale University Faculty of Arts and Sciences High Performance Computing Center.
\end{theacknowledgments}



\bibliographystyle{aipproc}   


\IfFileExists{\jobname.bbl}{}
 {\typeout{}
  \typeout{******************************************}
  \typeout{** Please run "bibtex \jobname" to optain}
  \typeout{** the bibliography and then re-run LaTeX}
  \typeout{** twice to fix the references!}
  \typeout{******************************************}
  \typeout{}
 }

\end{document}